\documentclass[reprint,aps,prd,nofootinbib,a4paper,10pt,twocolumn,superscriptaddress]{revtex4-2}
\usepackage{hyperref} 
\usepackage{changes}
\usepackage{graphicx,subfigure}
\usepackage[section]{placeins}
\usepackage[titletoc]{appendix}
\usepackage{xcolor}
\usepackage{dsfont}
\usepackage{bm}
\usepackage{mathtools} 
\usepackage{amsfonts}
\usepackage{enumerate}

\usepackage{dcolumn}
\usepackage{multirow}
\usepackage{enumitem}

\relax

\hypersetup{
     colorlinks   = true,
     citecolor    = cyan,
     linkcolor = pink
     }

\begin{document} 

\title{Diagnostic Consistency Tests of the Concordance Cosmology}

\author{S. M. Koksbang} 
\email{koksbang@cp3.sdu.dk}
\affiliation{CP$^3$-Origins, University of Southern Denmark, Campusvej 55, DK-5230 Odense M, Denmark}
\author{A. Heinesen}
\email{asta.heinesen@nbi.ku.dk}
\affiliation{Department of Physics and Astronomy, Queen Mary University of London, UK}
\affiliation{Niels Bohr Institute, Blegdamsvej 17, Copenhagen, 2100, Denmark}

\begin{abstract}
The $\Lambda$CDM cosmological model faces increasingly significant and robust tensions among independent cosmological probes, prompting renewed scrutiny of its foundational assumptions. While General Relativity and the nature of dark energy are now routinely tested with cosmological surveys, less progress has been made testing the space-time geometry at the largest scales, and in particular testing the assumption that observables (distances, redshifs, expansion of space, etc.) on the largest scales are described by a single Friedmann-Lema\^{\i}tre-Robertson-Walker (FLRW) metric.
In order to enable such tests, we introduce a model-independent framework that combines successive derivatives of the angular diameter distance, $d_A(z)$, with the line-of-sight expansion rate, $\mathcal{H}(z)$, to expose the physical content of well-known FLRW consistency relations. This allows us to perform diagnostic tests of the large-scale geometry, that are free of assumptions about dark energy and the theory of gravity on large scales. In addition, we derive a new nonparametric estimator for the cosmic density field that is independent of the Friedmann equations. This enables qualitatively new, observationally accessible tests of the FLRW framework and provides a stringent, model-independent diagnostic for departures from standard cosmology using current and forthcoming distance and expansion rate measurements.
\end{abstract}

\maketitle

\paragraph*{Introduction}
For decades, the $\Lambda$ Cold Dark Matter ($\Lambda$CDM) model has played the role of the concordance model in cosmology. However, as we have ventured into the era of precision cosmology, the $\Lambda$CDM model has run into difficulties with explaining observational data. Among the most prominent challenges is the Hubble tension (the disagreement between the Planck estimate of the Hubble constant, $H_0$, \cite{planck} and that obtained through the local distance ladder, e.g. \cite{local}), but recent analyses also highlight the difficulty of reconciling DESI, supernova, and Cosmic Microwave Background data within the $\Lambda$CDM framework \cite{DESI:2024mwx}.
\newline\indent
The growing cosmological tensions increasingly motivate testing the foundations of the $\Lambda$CDM framework itself, beyond simply fitting parameters to the model. Such tests are of fundamental importance for assessing the validity of the standard model for describing the Universe on large scales, and include tests of general relativity \cite{Ferreira:2019xrr,GRtest1, GRtest2, GRtest3, Thomas:2025qiz} and tests of the (often assumed flat) Friedmann-Lema\^{\i}tre-Robertson-Walker (FLRW) metric \cite{FLRWtest1, FLRWtest2, Montanari:2017yma, FLRWtest3, cbl, Larena:2008be, Seikel_2012}. 
The most well-known example of the latter type of test is the Clarkson-Basset-Lu (CBL) test \cite{cbl} 
\begin{align} 
\label{eq:C}
    \mathcal{C}(z) = 1 + H^2\left(DD'' - D'^2\right) + HH'DD',
\end{align}
where $D:=(1+z)d_A$, with $d_A$ the angular diameter distance to the source, and a prime denotes differentiation with respect to the redshift $z$. The practical use of this test is conducted by reconstructing the right-hand-side through observations. 
In any FLRW spacetime, the combination and thus $\mathcal{C} = \mathcal{C}(z)$ must equal zero at all redshifts and in all directions of observation. Thus, if independent observations of the angular diameter distance $d_A=d_A(z)$ (e.g. from supernovae) and $H=H(z)$ (e.g. from baryon acoustic oscillations, BAO, or cosmic chronometers \cite{Moresco:2022phi}) yield $\mathcal{C}\neq 0$, this indicates that the Universe is not well described by an FLRW model on the scales probed by the observational data. However, we do not {\em a priori} know which non-FLRW spacetimes break this condition and which do not. We also do not know what $\mathcal{C}$ expresses in general, i.e. for non-FLRW spacetimes. This is also true for other often-used FLRW consistency relations (see  a list and discussion in \cite{list}). Overall, these consistency relations are based on an approach that assumes FLRW geometry in the derivation of the test itself. 
As a result, while such tests can confirm internal consistency within the FLRW framework, they do not provide insight into how the same combinations of observables would behave in a more general spacetime. In non-FLRW models, the constructed quantities may deviate from the expected constant, but without a theoretical prediction for their behavior in such settings, the deviation cannot be straightforwardly interpreted. This limits the diagnostic power of these tests: it is currently unclear how their failure to yield the FLRW-expected result can be understood in alternative geometries. In this work, we overcome this limitation by providing derivatives of $d_A$ in a general spacetime. Working with $d_A, d_A'$, $d_A''$ as well as the measurable line-of-sight expansion $H$ and its derivative $H'$ as fundamental quantities, we can explicitly compute the geometrical expression for \eqref{eq:C}.  
\newline\indent
Instead of being an abstract number specifically tied to the FLRW assumptions, $\mathcal{C}$ is now a function defined on the observer's past lightcone which is described as a combination of kinematic and curvature variables that are well defined in a general spacetime. Furthermore, having access to the derivatives of $d_A$ in a general spacetime allows us to construct alternative combinations that are guaranteed to take simple, intuitive forms with clear physical meaning in \emph{any} spacetime. Thus, obtaining the expressions for $d_A, d_A'$, $d_A''$, $H$, and $H'$ in a general spacetime makes it easy to generate tests that are genuinely diagnostic: if the observed combinations deviate from the FLRW expectation, we can interpret the deviations in terms of physical quantities rather than simply as an abstract anomaly. 
\newline\indent
We may additionally construct new test statistics for the dynamics and energy-momentum content of the spacetime by imposing Einstein's field equation in statistics built from suitable combinations of distance measures (including their derivatives). In this spirit, we proceed to construct a combination that for perfect fluid general relativistic cosmologies yields the density field as a function of the redshift.

\begin{figure*}
    \centering
    \includegraphics[width=2\columnwidth]{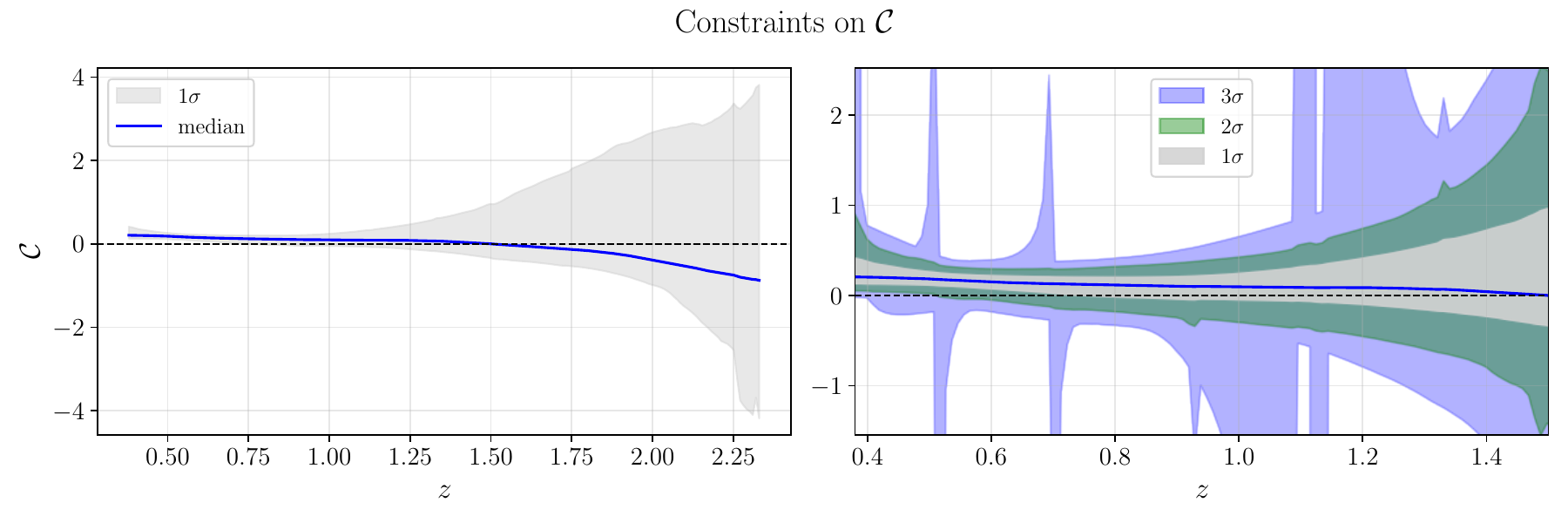}\\
    \includegraphics[width=2\columnwidth]{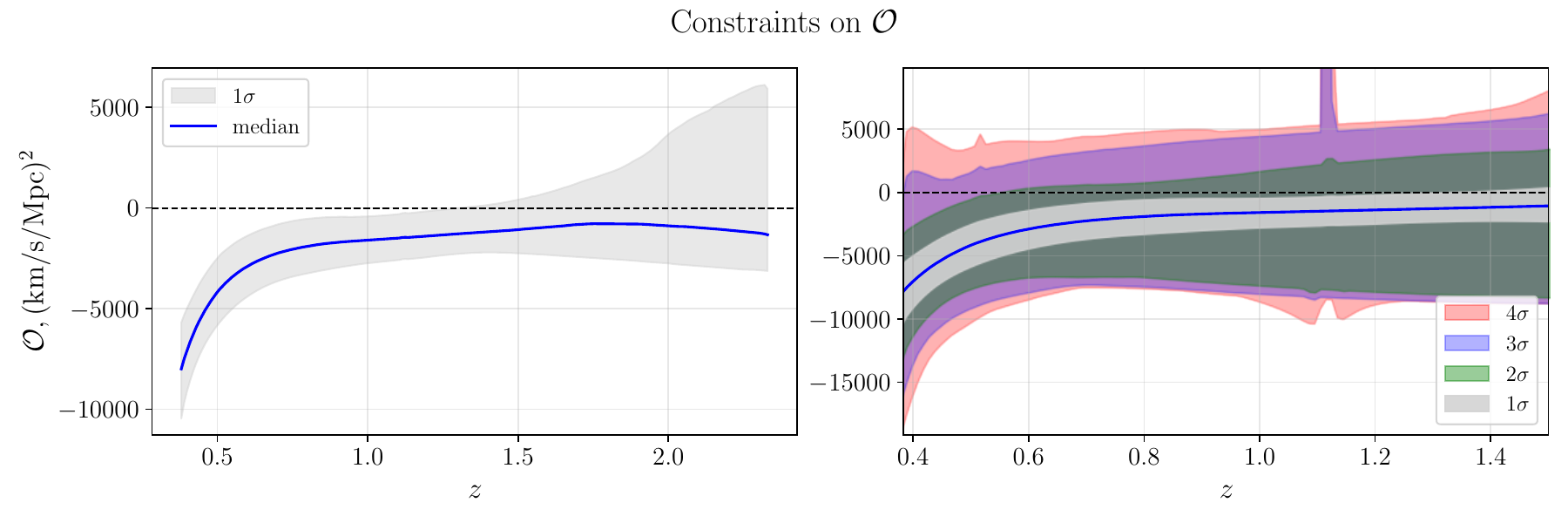}\\
    \includegraphics[width=2\columnwidth]{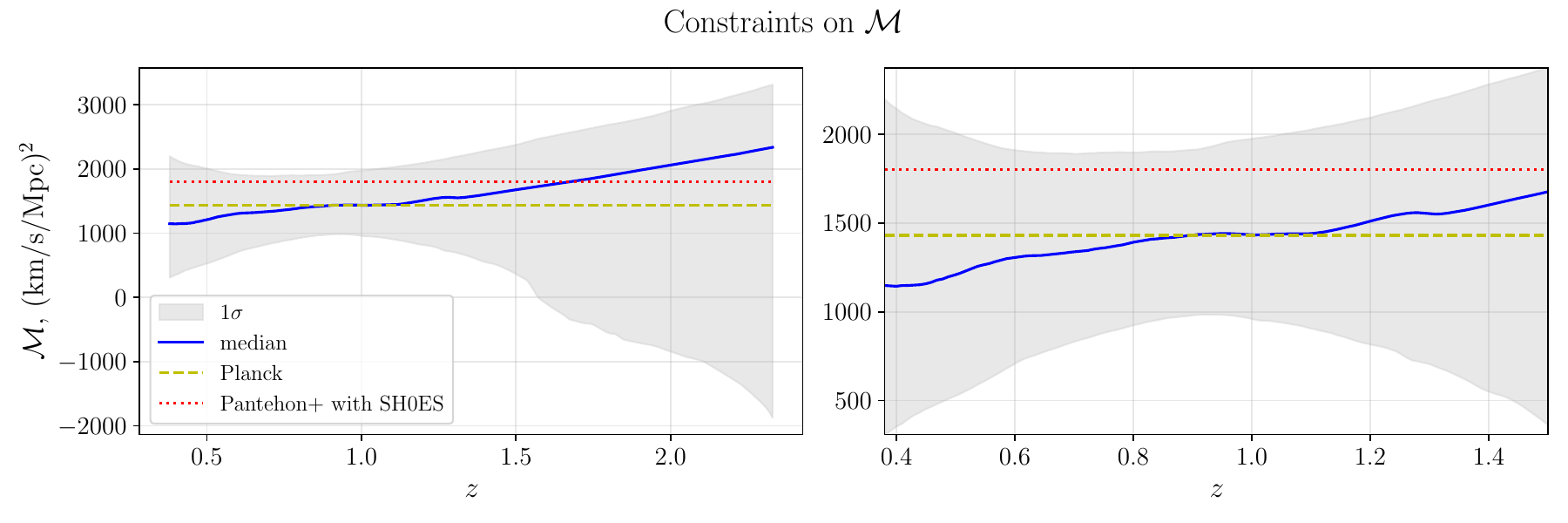}
    \caption{Constraints on $\mathcal{C}, \mathcal{O}$ and $\mathcal{M}$ in terms of median and percentiles that for Gaussian distribution would correspond to $\sigma$. $\mathcal{M}$ is shown together with its $\Lambda$CDM predictions based on best-fit Planck and SH0ES parameters. Dashed lines at 0 in the figures for $\mathcal{O}$ and $\mathcal{C}$ highlight the vanishing flat FLRW predictions. Close-ups are shown to the right. For the reader interested in comparing with \cite{PRD}, we note that the results were obtained using ``criteria set 3'' as described in \cite{PRD}. Note that spikes in the uncertainty bands appear because the considered symbolic expression reconstructions of $\mathcal{H}$ include expressions that diverge at certain redshifts. 
    }
    \label{fig:results}
\end{figure*}

\paragraph*{Diagnostic consistency relations}
In order to generalize known consistency relations involving the angular diameter distance and its derivative, we first need to derive the expressions for $d_A'$ and $d_A''$ for general spacetimes. We do this by utilizing Sachs optical equations as reparametrized in terms of redshift.
This has previously been done in the context of cosmography, cf., e.g., \cite{asta}, where expansions of $d_L$ and $d_A$ for a general spacetime was presented around the observer position at $z = 0$. Here, we augment the derivations of \cite{asta} to include $d_A'$ and $d_A''$ at arbitrary redshift. 
We first define the effective expansion rate $\mathcal{H}$
\begin{align}
    \mathcal{H} \, :=  \frac{\rm d (1/E)  }{{\rm d} \lambda  } = \frac{1}{3}\theta - e^\mu a_\mu + e^\mu e^\nu\sigma_{\mu\nu},
\end{align}
where $\lambda$ is the affine parameter along the light ray, increasing from the source to observer, and where $E$ is the photon energy as measured by observers that comoves with a hypothesized cosmic fluid description of observers in the space-time. In this paper, we shall work in units where the photon energy at the observer equals unity, such that the redshift function is simply given by $z = E - 1$.
In the above, $\theta$ is the local expansion rate of the cosmic fluid, $a^\mu$ its acceleration, $\sigma_{\mu\nu}$ its shear tensor and $e^\mu$ the spatial direction of emission. The generalized Hubble parameter, $\mathcal{H}$, naturally appears in quantifications of observables in general space-times (see e.g. \cite{Clarkson:2010uz} and its references) and reduces to the ordinary Hubble parameter $H(z)$ in the FLRW limit. $\mathcal{H}$ is what we effectively measure as the ``Hubble parameter'' in observations measuring the line-of-sight expansion (see e.g. \cite{asta}). It is explicitly the parameter obtained both from cosmic chronometers \cite{Heinesen:2024gdi} (averaged over a region around the particular galaxies used for the measurement) and the longitudinal BAO observations \cite{astabao1, astaboa2} (calibrated with the value of the sound horizon at the drag epoch). When averaging over many lines of sight in a perturbed FLRW universe, we expect that the acceleration and shear terms vanish so that $\mathcal{H}$ reduces to $1/3\cdot \theta$ which in the FLRW limit is again just the ordinary Hubble parameter.
\newline\newline
To derive $d_A'$ and $d_A''$, we first remember the relations \cite{Schneider1992} (chapters 3 and 4) and \cite{godbog}
\begin{align}
    \begin{split}
        \frac{\rm d}{{\rm d} z} & = -\frac{1}{(1+z)^2\mathcal{H}}\frac{{\rm d}}{ {\rm d} \lambda}\\
        \frac{{\rm d} \hat\theta}{ {\rm d} \lambda} &= -\frac{1}{2}\hat\theta^2 - 2|\hat\sigma|^2-R_{\mu\nu}k^\mu k^\nu\\
        \frac{{\rm d} d_A}{{\rm d} \lambda}& = \frac{1}{2}\hat\theta d_A,
    \end{split}
\end{align}
where $\hat\theta$ and $|\hat\sigma|^2$ are the optical expansion and shear scalars of the light beam. The first of these equations lets us switch between derivatives with respect to the affine parameter and the redshift, assuming that the function $z(\lambda)$ is invertible. The second equation is the transport equation for the optical expansion scalar, while the bottom equation relates the angular diameter distance to the optical expansion.
\newline\indent
After differentiating $d_A$ and rewriting the result for $d_A''$ we obtain
\begin{align}\label{eq:dadaprime}
\begin{split}
    d_A' &= -\frac{\hat \theta}{2(1+z)^2\mathcal{H}}d_A\\
    d_A''  & =\frac{d_A}{2(1+z)^4\mathcal{H}^2}\times \\& \left( -2|\hat\sigma|^2 - R_{\mu\nu}k^\mu k^\nu + 2\hat\theta(1+z)\mathcal{H}+ (1+z)^2\hat\theta\mathcal{H}'  \right).
\end{split}
\end{align}
With these expressions at hand, we can readily write up the standard FLRW consistency relations and see what they yield in a general spacetime. For the CBL test we find that the theoretical interpretation of the combination on the right-hand-side of eq. \ref{eq:C} can be expressed as
\begin{align}\label{eq:Cderived}
    \begin{split}
    \mathcal{C}&= 1+ \frac{d_A^2}{(1+z)^2}\left(-|\hat\sigma|^2-\frac{1}{2}R_{\mu\nu}k^\mu k^\nu + \hat\theta(1+z)\mathcal{H}-\frac{\hat\theta^2}{4}\right)\\& + d_A^2\left( \mathcal{H}\mathcal{H}'(1+z)-\mathcal{H}^2 \right) ,
    \end{split}
\end{align}
valid for a general spacetime where the geometrical optics approximation applies for light/gravitational waves\footnote{We remind the reader that the geometric optics approximation describes the propagation of waves in the limit where their wavelength is much smaller than the characteristic length scales of the background. In this limit, waves travel along null geodesics. %well‑defined rays (null geodesics) with slowly varying amplitude, and diffraction and interference effects are neglected so that the wave dynamics reduce to ray tracing governed by the local geometry.
}. 
Since $\mathcal{H}$ and the other expansion and curvature variables entering in \eqref{eq:Cderived} are non-isotropic, $\mathcal{C}$ is generally not isotropic and thus has dependence on the direction of observation as well as on the affine distance to the source.
\newline\indent
For a general spacetime, the expression on the second line in Eq. \eqref{eq:Cderived} does not reduce any further and we thus conclude that the CBL statistic is composed of several terms encoding expansion and deformation of the null geodesic bundle and expansion degrees of freedom in the frame of the observers in the spacetime.   
We might naively have expected that the CBL test would be a probe of the spatial curvature of the spacetime, since the reason that the  CBL test is zero in the FLRW geometry is that it reduces to the derivative of the constant FLRW curvature parameter. However, we see that \eqref{eq:Cderived} generally does not have a simple interpretation in terms of spatial curvature.
\newline\newline
Because constraints on derivatives are weaker than those directly on $d_A$ and $\mathcal{H}$ and higher-order derivatives even less so, we also consider the $z-$integral of the CBL test, $\mathcal{O}$, which we, using the relation in Eq. \ref{eq:dadaprime}, find to be
\begin{align}\label{eq:O}
\begin{split}
    \mathcal{O} &= \frac{\mathcal{H}^2D'^2-1}{D^2} \\&= \frac{\mathcal{H}^2\left( 1+\frac{\hat\theta^2}{4(1+z)\mathcal{H}^2} - \frac{\hat\theta}{(1+z)\mathcal{H}} \right)-1/d_A^2}{(1+z)^2}
    \\&\underset{\mathclap{\text{FLRW}}}{=} \,\,\,\,\,\Omega_{k,0}H_0^2,
\end{split}
\end{align}
where $\Omega_{k,0}$ is the FLRW curvature parameter and where the top line presents the observable combination while the following lines demonstrate the theoretical interpretations of the combination. In a general spacetime, the expression on the second line cannot be simplified any further. However, in the FLRW limit, it reduces to a simple factor of the curvature parameter and squared Hubble constant.
\\\\
We now explore alternative combinations of $d_A$ and its derivatives to construct a combination of $d_A, d_A'$, $d_A''$, $\mathcal{H}$ and $\mathcal{H}'$ that remain transparent and physically interpretable beyond FLRW cosmology. One particularly interesting example of a combination of the above quantities is
\begin{align}\label{eq:density}
\begin{split}
\mathcal{M}:&=\,\,\,\,\,\,\,\,\,\,\,\,-\frac{2\mathcal{H}^2}{3d_A(1+z)}\left(d_A'\left[\frac{2}{1+z}+\frac{\mathcal{H}'}{\mathcal{H}}\right] + d_A''\right)\\&= \frac{2|\hat\sigma|^2+R_{\mu\nu}k^\mu k^\nu}{3(1+z)^5 }\\ &\underset{\mathclap{\text{perfect fluid}}}{=}\,\,\,\,\,\,\,\,\,\,\,\,\frac{2|\hat\sigma|^2}{3(1+z)^5 } + \frac{8\pi G}{3(1+z)^3}(\rho+p)\\ &\underset{\mathclap{\text{$\Lambda$CDM}}}{=} \,\,\,\,\,\,\,\,\,\,\,\,\Omega_{m,0}H_0^2.
\end{split}
\end{align}
The right-hand-side of the top line presents the combination of observables defining $\mathcal{M}$ while the following lines illustrate the theoretical interpretation of this combination. In the perfect fluid limit and assuming that general relativity holds, $R_{\mu\nu}k^\mu k^\nu$ reduces to $R_{\mu\nu}k^\mu k^\nu = 8\pi G(\rho + p)(1+z)^2$. 
We thus see that the combination on the left-hand-side provides the total density field (assuming a rigid equation of state $p\propto \rho)$ in general spacetimes on scales where we can treat the cosmological content as a perfect fluid with negligible Weyl focusing through optical shear. Furthermore, since $|\hat\sigma|^2\ge 0$, a negative sign of the combination on the left-hand-side would indicate a violation of the null-energy-condition which requires $R_{\mu\nu}k^\mu k^\nu$ to remain strictly non-negative. 
In the matter and cosmological constant dominated era of the $\Lambda$CDM model, this simplifies even further, permitting us to write the right-hand-side of the equation as simply the constant number $\Omega_{m,0}H_0^2$. 
Thus, \eqref{eq:density} provides both a useful model-independent probe of cosmic density and a new $\Lambda$CDM consistency test.
\\\\
\paragraph*{Constraints on $\mathcal{C}, \mathcal{O}$ and $\mathcal{M}$}
In a companion paper, \cite{PRD}, we propose a model-independent bootstrap-based symbolic regression approach for constraining $\mathcal{C}, \mathcal{O}$ and $\mathcal{M}$. Symbolic regression makes it possible to constrain quantities such as $d_A, \mathcal{H}$ and their derivatives without assuming a cosmological model. In bootstrap-based symbolic regression, symbolic regression is performed multiple times over different bootstraps of data sets in order to obtain estimates of the uncertainty of the resulting constraints. To obtain uncertainty bands, we have thus used 200 different symbolic expressions for $d_A$ and $\mathcal{H}$ each, combining them in all $4\cdot 10^4$ unique combinations when constraining $\mathcal{C}, \mathcal{O}$ and $\mathcal{M}$. From these combinations, we calculate median and percentages corresponding to standard deviations for Gaussian distributions. We refer the reader to \cite{PRD} for details on the approach and considerations on the robustness of the obtained results. Here, we merely show the main constrains obtained using the Pantheon+ \cite{pantheon} data set to constrain $d_A, d_A'$ and $d_A''$, and DESI data release 2 \cite{DR2} (table IV) together with a single BOSS data point from \cite{BOSS72} (the $z = 0.38$ data point from table 7) to constrain $\mathcal{H}$. 
%In the original versions of $\mathcal{C}$ and $\mathcal{O}$, there is an algebraic cancellation of the sound horizon. This algebraic cancellation vanishes in the diagnostic generalizations we have presented. When constraining $\mathcal{H}$ 
We marginalize over the sound horizon, sampling over deviations $\delta r_s\in[-2, 2]$ (using uniform priors) from the Planck value $146.995$Mpc. When constraining $d_A$ and its derivatives, we use the reported distance moduli rather than the raw data since the raw-data parameters $\alpha, \beta$ and $\Delta_M$ do not vary notably between various test cosmologies \cite{Dam:2017xqs}. The results are shown in figure \ref{fig:results} where we see that current data cannot be used to constrain $\mathcal{M}$ very tightly, and the $\Lambda$CDM prediction with both Planck and Pantheon parameter values lies within $1\sigma$ om the median result. Both $\mathcal{C}$ and $\mathcal{O}$ show deviation from the flat FLRW assumption at up to $2-3\sigma$. As we discuss in \cite{PRD}, our bootstrap-based symbolic regression cannot be considered fully robust in terms of quantifying uncertainty and we here show those of our DESI data release 2 results that show the mildest violation of FLRW consistency. As discussed and studied in \cite{PRD}, the exact quantitative uncertainty bands depend on the hyperparameters of the used symbolic regression algorithms as well as the criteria used for retaining/rejecting expressions found by the algorithms. For the results presented here, we used an automated set of criteria for rejecting/retaining expressions reconstructing $\mathcal{H}$ (``criteria set 3'' in \cite{PRD}). This criteria set was based on a calculation of the expressions' complexity and precision and did not remove all pathological expressions such as those undefined or divergent on parts of the considered redshift interval. Their appearance among our 200 symbolic expressions for $\mathcal{H}$ leads to broader uncertainty bands than if pathological expressions are removed, and also leads to spikes in the uncertainty bands (clearly seen in figure \ref{fig:results}).

\paragraph*{Conclusion}
In this work, we have expressed the Clarkson–Bassett–Lu (CBL) test and its integral in fully general spacetimes, thereby converting these long‑standing FLRW consistency tests into genuinely interpretable geometric diagnostics. By deriving $d_A'$, $d_A''$, $\mathcal{H}$, and $\mathcal{H}'$ in a model‑independent way, we identify precisely which kinematic and curvature quantities govern deviations from the FLRW prediction, allowing any observational violation to be traced to physically meaningful properties of the spacetime rather than treated as an abstract inconsistency. Our results are valid for any metric theory of gravity, assuming that the geometric optics approximation holds so that light travels along null-geodesics, and assuming that the affine parameter is related invertibly to the redshift so that we can convert from $d/d\lambda$ to $d/dz$.
Beyond clarifying the foundations of existing tests, we have constructed a new combination of observables, $\mathcal{M}$, which yields the total density field for any perfect‑fluid cosmology if we further assume general relativity to hold, but still independent of the Friedmann equations or FLRW symmetry assumptions. In the appropriate limits this reduces to a direct, model‑independent probe of $\Omega_{m,0} H_0^2$, while in more general settings it constrains null‑energy‑condition violations and Weyl focusing. This diagnostic therefore extends earlier proposals by providing a single, observationally accessible quantity with a transparent physical interpretation across arbitrary geometries.
We demonstrate the possibility of constraining the generalized tests as well as $\mathcal{M}$ with current data in a companion paper, \cite{PRD}, and show some of the main results here. These results show a $\sim 2-3\sigma$ tension with the FLRW assumption, but we caution that further study and possibly refinement of our method for constraining the quantities is required before we can be certain of the robustness of the results. Nonetheless, it is worth noting that we consistently find FLRW violation at $2\sigma$ level or above when we modify subjective choices of our bootstrap-based symbolic regression methodology such as the criteria used to retain or discard expressions (e.g. the tolerance for pathological behavior and goodness-of-fit thresholds when reconstructing $d_A, \mathcal{H}$ and their derivatives). We also note that our constraints on $\mathcal{O}$ and $\mathcal{C}$ correspond to a calibration offset of $D$ relative to $1/\mathcal{H}$ of $-10\%$ as discussed in \cite{PRD}. This is exactly the calibration offset between Pantheon+ and the BAO data when the latter is calibrated with the CMB data. This may also explain why we find the strongest apparent violation of FLRW at the lowest redshifts; a calibration offset is expected to have the largest impact in this regime, whereas at higher redshifts its effect may be partially diluted due to the integrated nature of distance measures along the lightcone.
\newline\indent
The current bottleneck of our methodology is the reconstruction of $\mathcal{H}$ and its derivatives. Supernova data is already ample and precise enough to permit precise reconstructions of $d_A$ and its derivatives. Fortunately, the coming decade will present us with not only much more supernovae data (such as from LSST), but also more BAO and cosmic chronometers data from e.g. Euclid which we expect will significantly improve the constraining power of our methodology. As precision measurements of distances and expansion rates continue to improve, our analytical results supply the theoretical framework needed to transform geometric observables into robust, model‑independent tests of the large‑scale structure of spacetime. This opens the door to a qualitatively new class of consistency relations: tests that not only detect departures from FLRW behavior but also reveal their physical origin. We lastly note that the main significance of the presented framework might not lie in the consistency tests themselves, but rather in the demonstration that cosmological observations can be constructed and analyzed from first principles without assuming phenomenological models. We look forward to seeing developments and implementations of new cosmological tests based on first principles in the future.

\begin{acknowledgments}
We thank Chris Clarkson for comments on our draft.
SMK is funded by Villum Fonden, grant VIL53032. 
AH is supported by the Perren Fund at the University of London, and wishes to thank the Astronomy Unit at Queen Mary University of London for their support.
{\bf Author contribution statement}: The analytical derivations were obtained by SMK and independently verified by AH. The numerical work behind the figures was conducted by SMK who led the overall development of the project. Both authors contributed substantially to the refinement of the work and to the writing of the manuscript.
\end{acknowledgments}

\bibliography{bibliography}

\end{document}